\documentclass[a4paper,11pt]{article}
\usepackage{submission}
\usepackage{graphicx}
\usepackage{color}
\usepackage{hyperref}

\title{\vspace*{1cm}\boldmath Determination of the sensitivity of $\Lambda$ and $\Lambda^+_c$ electric dipole moments using a full angular analysis}
\author[a]{R.~T.~Ovsiannikov,}
\author[a,b,c]{A.~Yu.~Korchin,}
\author[d]{E.~Kou}
\affiliation[a]{NSC Kharkiv Institute of Physics and Technology, 61108 Kharkiv, Ukraine}
\affiliation[b]{V.N.~Karazin Kharkiv National University, 61022 Kharkiv, Ukraine}
\affiliation[c]{M.~Smoluchowski Institute of Physics, Jagiellonian University, 30-348 Krakow, Poland}
\affiliation[d]{Universite Paris-Saclay, CNRS/IN2P3, IJCLab, 91405 Orsay, France}

\emailAdd{roman.ovsiannikov@kipt.kharkov.ua}
\emailAdd{korchin@kipt.kharkov.ua}
\emailAdd{emi.kou@ijclab.in2p3.fr}
\date{\today}

\abstract{
A search for a non-zero Electric Dipole Moment (EDM) of particles, which is a clear signal of a violation of CP symmetry,  is one of the unique ways to discover physics beyond the Standard Model. In this paper, we discuss a method for determining the EDM of baryons from the full angular distribution of final particles in electron-positron pair annihilation processes. The question of how accurately  the state-of-the-art experiments can determine  EDM 
of $\Lambda$ and $\Lambda_c^+$ baryons is discussed in detail. Investigating the pseudo-statistics that corresponds to the BESIII experiment, the estimated sensitivity for $\Lambda$ EDM is obtained at the level of $10^{-18}$ $e\,$cm.  The similar figure for the proposed Super Tau-Charm Facility (STCF) experiment is found to be of the order 
of $10^{-20}$ $e\,$cm. For the $\Lambda_c^+$ EDM, the calculated sensitivity for the STCF experiment is $10^{-16}$ $e\,$cm. The case of a polarized initial electron is considered separately as such an option is planned 
at the STCF experiment.}

\begin{document}

\maketitle

\setcounter{footnote}{0}

\newpage
\section{Introduction}

The search for electric dipole moments (EDMs) of elementary particles has been going on for many years. Observing a nonzero EDM of an elementary particle means a simultaneous breaking of P and CP symmetries. 
There are two sources of CP symmetry breaking in the Standard Model (SM). One is the unremovable 
complex phase $\delta_{KM}$ in the Cabibbo-Kobayashi-Maskawa (CKM) matrix. Another source is a possible term in the QCD Lagrangian, the so-called $\theta$ term. Although $\theta$ is unknown, the observed restriction on the 
neutron EDM suggests that $\theta < 10^{-(9\pm 1)}$ \cite{Bigi:2000yz}. Since the natural scale is of 
$\theta \sim O(1)$, the very small value of $\theta$ (known as the strong CP problem) requires an explanation. 

 In this article, we investigate a possible measurement of the hyperonic EDMs.
{Currently, the experimental constraint on the hyperon EDM is rather weak:  
 $d_\Lambda < 1.5\times 10^{-16}$ $e\,$cm, CL=95\%~\cite{ParticleDataGroup:2022pth}.    
This value is very large compared to the most stringent constraint for baryons measured for the neutron, 
$d_n< 1.8 \times 10^{-26}$ $e\,$cm, CL=90\%~\cite{Abel:2020pzs}. } 
 For the EDM of $\Lambda^+_c$,  there are no direct experimental  constraints at all as of today.  
On the other hand, there is a new experimental proposal to measure the $\Lambda^+_c$ EDM using bent crystals, which would achieve an EDM constraint of the order of $10^{-16}$ $e\,$cm~\cite{Aiola:2020yam}. That proposal is based on phenomenon of the baryon spin precession in strong field of a bent crystal when the baryon  
moves in the channeling regime in the crystal~\cite{Botella:2016ksl, Fomin:2017ltw, Bagli:2017foe, 
Fomin:2019wuw}.  
In the present article we discuss an alternative method to obtain the EDMs of $\Lambda$ and $\Lambda^+_c$ hyperons by analyzing their decay products.  

A large number of $\Lambda$ hyperon-antihyperon pairs can be easily produced from $J/\psi$ decays at tau-charm factories. For example, the BESIII experiment is capable of generating $10^{10}$ $J/\psi$  that produce $16 \times 10^6$ $\Lambda \bar{\Lambda}$ pairs per year \cite{Li:2016tlt}.  The $\Lambda \bar{\Lambda}$ and $\Lambda_c^+ \bar{\Lambda}_c^-$ pairs can also be produced directly from the electron-positron annihilation  to a photon. 
The STCF project is capable of producing $5.6 \times 10^8$ $\Lambda_c^+ \bar{\Lambda}_c^-$ pairs \cite{Achasov:2023gey}. The identification of hyperon-antihyperon pairs is usually performed with very high accuracy, exceeding $95\%$ \cite{Achasov:2023gey}. 
 The process of hyperon-antihyperon pair production in electron-positron annihilation via an intermediate photon or $J/\psi$ allow to measure the cross section, thus the Electric Dipole (ED) form-factor. Then, the EDM can be obtained by an analytical continuation of the form-factor from the region of nonzero squared 
four-momentum transfer, $q^2 \ne 0$, to $q^2=0$. 
  
The present paper investigates a method for estimating EDM from the full {angular} analysis for the following processes: $e^+ e^- \rightarrow J/\psi \rightarrow \Lambda \bar{\Lambda} \rightarrow p\pi^- \bar{p} \pi^+$ and $e^+ e^- \rightarrow \Lambda_c^+ \bar{\Lambda}_c^- \rightarrow \Lambda\pi^+ \bar{\Lambda} \pi^-$. These processes are analyzed with the  statistics estimated for the BESIII and STCF experiments; the polarization of the initial electron is  considered, and the sensitivity of the EDM depending on this parameter is also investigated. For the former of the above processes, a full angular analysis has already been performed 
\cite{BESIII:2018cnd, BESIII:2022qax}, but these studies did not consider EDM effects. 
{For the $\Lambda_c^+$, the analysis of $e^+ e^- \rightarrow \Lambda_c^+ \bar{\Lambda}_c^-$ was carried out without taking into account the baryon decay, also without considering EDM terms~\cite{BESIII:2023rwv}.} In our work, we extend these analyses  by including the effects associated with EDMs. 

Note that for the EDM of $\Lambda$,  somewhat similar method which uses 
formalism of helicity amplitudes was recently applied in Ref.~\cite{Fu:2023ose}. In our paper, 
direct calculation of the cross section for polarized hyperon and antihyperon is carried out 
for both cases of polarized and unpolarized electron beams. 
The computational method follows those developed widely for unstable polarised fermion-antifermion pair production 
in the $e^+ e^-$ annihilation process. Indeed, such processes attracted considerable attention in the literature, in particular, the study of CP violation and anomalous couplings in 
the $t \bar{t}$ production~\cite{Bernreuther:1992dz, Bernreuther:1992be, Arens:1992wh, Arens:1994jp, Grzadkowski:1996pc, Grzadkowski:2000nx, Truten:2021rrq} or the production of $\tau^+ \tau^-$ pairs~\cite{Bernreuther:1993nd, Bernabeu:2007rr, Bernabeu:2006wf, Bernabeu:2008ii, 
Bernreuther:2021elu, Bernreuther:2021uqm, Banerjee:2022sgf, Banerjee:2023qjc}. Of course this list of the references is far from being complete. We should also mention the study \cite{Czyz:2007wi} of the $\Lambda \bar{\Lambda}$ production in which the accent was placed on the radiative return method and its implementation in the Monte Carlo generator PHOKHARA, although  the EDM effects were not included.  

The remainder of the paper is organized as follows. In Section~\ref{sec:theory} theoretical formalism for calculation of the cross section with polarized hyperons is presented. Section~\ref{sec:pseudo-data} describes procedure of generating 
pseudo-data which are used in the analysis. The results on the sensitivity study of EDM measurements are given in Section~\ref{sec:results} and we conclude in  
Section~\ref{sec:conclusions}. Finally, Appendix~A lists elements of the spin-correlation matrix used in calculation of the cross section in Section~\ref{sec:theory}.

\section{Theoretical framework}
\label{sec:theory}

Let us write down the Lorentz-invariant matrix element of the process $e^+ e^- \rightarrow \Lambda \bar{\Lambda}
\, (\Lambda_c^+ \bar{\Lambda}_c^-)$:
\begin{eqnarray}\label{eq1}
    \mathcal{M} &=& \frac{e^2}{s} \bar{v}(k_2) \gamma_\mu u(k_1)  \nonumber\\ 
    &\times&  \bar{u}(p_1)\left\{G_M(s) \gamma^\mu  + \frac{(p_2 - p_1)^\mu}{2M}\Big(\frac{G_E(s) - G_M(s)}{\gamma^2 - 1} - i\gamma_5 F_3(s)\Big)\right\} v(p_2),
\end{eqnarray}
 {where} $G_E(s)$ ($G_M(s)$) is electric (magnetic) form-factor and $F_3(s)$ is ED form-factor, where 
$s=(k_1+k_2)^2=(p_1+p_2)^2$ and $\gamma=\sqrt{s}/(2 M)$ is 
the Lorentz factor. 
Also here $M$ is the mass of the baryon $\Lambda$ ($\Lambda_c^+$), $k_1$ ($k_2$) is the four-momentum of the electron (positron), $p_1$ ($p_2$) is the four-momentum of the baryon (antibaryon). 

The cross-section of the process   $e^+ e^- \rightarrow \Lambda \bar{\Lambda}
\, (\Lambda_c^+ \bar{\Lambda}_c^-)$ is
\begin{equation}
\label{eq2}
    \dfrac{d \sigma (e^+ e^- \rightarrow  \Lambda \bar{\Lambda} \, (\Lambda_c^+ \bar{\Lambda}_c^-))}{d \Omega}  = \dfrac{\beta}{64 \pi^2 s} \overline{|\mathcal{M}|^2},
\end{equation}
where $\beta = \sqrt{1-4 {M^2}/{s}}$ is the baryon velocity. 

To obtain the cross-section of the two-step process $e^+ e^- \rightarrow  \Lambda \bar{\Lambda} \rightarrow p\pi^- \bar{p} \pi^+$ ($e^+ e^- \rightarrow \Lambda_c^+ \bar{\Lambda}_c^- \rightarrow \Lambda\pi^+ \bar{\Lambda} \pi^-$), we introduce polarization of the  intermediate baryons 
and denote  $\vec{s}$  {as} the polarization of the $\Lambda$ ($\Lambda_c^+$), 
$\vec{s}^{\, \prime}$  {as} the polarization of the $\bar{\Lambda}$ ($\bar{\Lambda}_c^-$), defined in their corresponding rest frames. 
The Cartesian components of the vectors $\vec{s}=(s_x, \, s_y, \, s_z)$ and 
$\vec{s}^{\,\prime}= (s'_x, \,  s'_y, \,  s'_z)$ are defined in the frame with OZ axis along the momentum $\vec{p}_1$ of baryon $\Lambda$ ($\Lambda_c^+$),  the plane XZ 
is spanned on the momenta of electron $\vec{k}_1$ and $\Lambda$ ($\Lambda_c^+$) baryon $\vec{p}_1$, 
and OY axis is along $\vec{p}_1 \times \vec{k}_1$.    

We also include the polarization of the electron. For the ultra-relativistic electron one needs only the longitudinal component with degree of polarization $\lambda_e$ ($|\lambda_e| \leq 1$). 

\begin{figure}[h]  
\centerline{\includegraphics[width=0.75\linewidth]{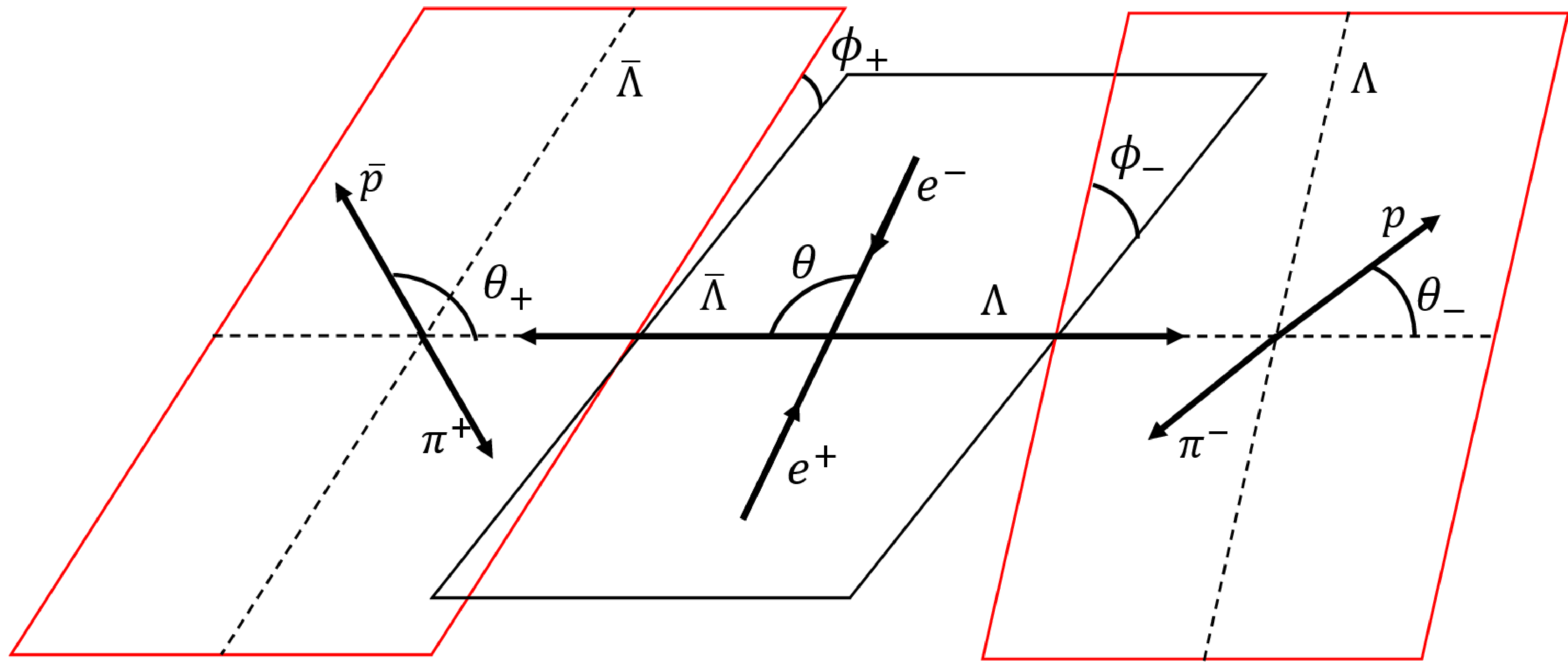}}
\caption{ The kinematical variables for the processes, $e^+ e^- \rightarrow  \Lambda \bar{\Lambda} \rightarrow p\pi^- \bar{p} \pi^+$.}
\label{ris:f0}
\end{figure}

We consider the two-body nonleptonic decays of $\Lambda$ and $\Lambda_c^+$ ($1/2^+ 
\rightarrow 1/2^+ 0^-$). The cross-section of the two-step process $e^+ e^- \rightarrow  \Lambda \bar{\Lambda} \rightarrow p\pi^- \bar{p} \pi^+$ can be written as \cite{Tsai:1971vv, Kawasaki:1973hf}
\begin{eqnarray}
    \frac{d \sigma (e^+ e^- \rightarrow  \Lambda \bar{\Lambda} \rightarrow p\pi^- \bar{p} \pi^+)}{d \Omega 
		 d\Omega_- d\Omega_+}
    &=&  \frac{d \sigma (e^+ e^- \rightarrow  \Lambda \bar{\Lambda})}{d \Omega } 
		\Bigg|_{{ \vec{s} \rightarrow \alpha_- \vec{n}, \; \vec{s}^{\, \prime} 
		\rightarrow \alpha_+ \vec{n}^{\, \prime} }} \nonumber\\ 
     &\times& \frac{1}{ 4\pi^2} {\rm Br}(\Lambda \rightarrow p\pi^-) 
		{\rm Br}(\bar{\Lambda} \rightarrow \bar{p} \pi^+), 
		\label{eq3}
\end{eqnarray}
\begin{eqnarray}
\label{eq:n,n'}
&& \vec{n} = (\sin \theta_- \cos \phi_-, \,  \sin \theta_-  \sin \phi_-, \, \cos \theta_-), \\
&& \vec{n}^{\, \prime} = (\sin \theta_+ \cos \phi_+, \,  \sin \theta_+ \sin \phi_+, \,  \cos \theta_+),  \nonumber
\end{eqnarray}
where the angles $\theta_-$, $\phi_-$ and $\theta_+$, $\phi_+$ (see Fig.~\ref{ris:f0}) describe 
direction of momenta of the proton and antiproton in the rest frame of $\Lambda$ and $\bar{\Lambda}$, 
respectively. 
Further, $d \Omega = \sin \theta d \theta d\phi$ with $\theta$ being the angle between the  
 momenta of $e^-$ and $\Lambda$ in the center-of-mass system (cms), and 
$\alpha_-$ and $\alpha_+$ are asymmetry parameters (polarization analyzers) in the 
$\Lambda$ and $\bar{\Lambda}$ decays~\cite{ParticleDataGroup:2022pth}. 
The latter are defined through the decay width of a hyperon in the rest frame:
\begin{eqnarray}
\label{eq:Gamma}
&&\frac{1}{\Gamma}\frac{d \Gamma (\Lambda \to p \pi^-)}{d \Omega} 
= \frac{1}{4 \pi} \bigl(1 +  \alpha_- \vec{s} \, \vec{n} \bigr),  \\
&&\frac{1}{\Gamma}\frac{d \Gamma (\bar{\Lambda} \to \bar{p} \pi^+)}{d \Omega} = 
\frac{1}{4 \pi} \bigl(1 + \alpha_+ \vec{s}^{\, \prime} \, \vec{n}^{\, \prime} \bigr), \nonumber 
\end{eqnarray}
where $\Gamma$ is the width of the $\Lambda$ ($\bar{\Lambda}$) decay into the channel 
$ p \pi^-$ ($\bar{p} \pi^+$). 

For the process $e^+ e^- \rightarrow \Lambda_c^+ \bar{\Lambda}_c^- \rightarrow \Lambda\pi^+ \bar{\Lambda} \pi^-$ the description is similar, only the meaning of indices ``+'' and ``-'' is reversed now because they refer to 
the charge of the final $\pi$-meson. Therefore, instead of $\alpha_-$ and $\alpha_+$, we use in this case the notation 
$\alpha$ for decay of $\Lambda_c^+$ and $\bar{\alpha}$ for decay of $\bar{\Lambda}_c^-$. 

One can decompose the cross-section of the process with a polarized  electron and polarized  baryons as 
\begin{eqnarray}\label{eq4}
    \dfrac{d\sigma(e^+ e^- \rightarrow  \Lambda \bar{\Lambda})}{d\Omega} &=& 
		\dfrac{d\sigma (e^+ e^- \rightarrow  \Lambda \bar{\Lambda})}{d\Omega}\Bigg{|}_{no \, EDM} 
		+ \dfrac{d\sigma (e^+ e^- \rightarrow  \Lambda \bar{\Lambda})}{d\Omega}\Bigg{|}_{EDM} \\
   && + \dfrac{d\sigma (e^+ e^- \rightarrow  \Lambda \bar{\Lambda})}{d\Omega}\Bigg{|}_{\lambda_e}. 
\end{eqnarray}
 In the following each of the contributions to the cross-section  (\ref{eq4}) is considered.
To simplify the expressions we separate the phases in the complex form factors
\begin{equation}
\label{eq:GE_GM}
G_E = |G_E| e^{i\Phi_E}, \quad G_M = |G_M| e^{i\Phi_M}, \quad F_3 = |F_3| e^{i\Phi_3},
\end{equation}
and introduce phase differences $\Delta \Phi_{EM} \equiv \Phi_E - \Phi_M $ and
$\Delta \Phi_{M3} \equiv \Phi_M -\Phi_3 $. 

It is also convenient to define 
\begin{equation} 
\label{eq:Sigma_psi}
\Sigma_\psi =  \gamma^2 |G_M|^2 +  |G_E|^2,
\end{equation}
and parameters $\alpha_\psi$ and $f_\psi$:  
\begin{equation}\label{eq7}
\alpha_\psi = \dfrac{\gamma^2 |G_M|^2 - |G_E|^2 }{\Sigma_\psi}, \qquad
f_\psi = \dfrac{\sqrt{2}|F_3|\gamma^2 \beta}{\sqrt{\Sigma_\psi}}.
\end{equation}
 
 Let us begin with the cross-section without EDM and electron polarization
\begin{eqnarray}
\dfrac{d\sigma (e^+ e^- \rightarrow  \Lambda \bar{\Lambda})}{d\Omega}\Bigg{|}_{no\,EDM} &=& \dfrac{\beta e^4 \Sigma_\psi}{2^{10} \pi^2 \gamma^4 M^2} \Big\{1 + \alpha_\psi \cos^2 \theta +
\sin^2 \theta (s_x s'_x - \alpha_\psi s_y s'_y)\nonumber \\  
&+&(\alpha_\psi + \cos^2 \theta) s_z s'_z + \cos \Delta \Phi_{EM} \sin \theta \cos \theta \sqrt{1 - \alpha_\psi^2} (s_x s'_z + s_z s'_x) \nonumber \\  
&+& \sin \Delta \Phi_{EM} \sin \theta \cos \theta \sqrt{1 - \alpha_\psi^2} (s_y + s'_y)\Big\},
\label{eq:CS_no_EDM_polarization}
\end{eqnarray}

Next, consider the EDM term, in which we keep only the terms linear in $f_\psi$  
as one can expect that $|F_3| \ll |G_E|, |G_M|$. It has the form
\begin{eqnarray}
\dfrac{d\sigma (e^+ e^- \rightarrow  \Lambda \bar{\Lambda})}{d\Omega}\Bigg{|}_{EDM} 
&=& f_\psi  \dfrac{\beta e^4 \Sigma_\psi}{2^{10} \pi^2 \gamma^4 M^2} \Big\{\cos (\Delta \Phi_{EM} + \Delta \Phi_{M3}) \sin^2 \theta \sqrt{1 - \alpha_\psi} \nonumber  \\
&\times&  (s_x s'_y - s_y s'_x) - \cos \Delta \Phi_{M3} \sin \theta \cos \theta \sqrt{1 + \alpha_\psi}(s_y s'_z - s_z s'_y)
\nonumber \\
 &-& \sin \Delta \Phi_{M3} \sin \theta \cos \theta \sqrt{1 + \alpha_\psi} (s_x - s'_x) \nonumber\\
&+& \sin (\Delta \Phi_{EM} + \Delta \Phi_{M3}) \sin^2 \theta \sqrt{1 - \alpha_\psi}(s_z - s'_z) \Big\}.
\label{eq:CS_with_EDM}
\end{eqnarray}
 Here the polarization of the initial electron is not included. 

 Adding the terms with the electron  polarization, we obtain
\begin{eqnarray}
\dfrac{d\sigma (e^+ e^- \rightarrow  \Lambda \bar{\Lambda})}{d\Omega}\Bigg{|}_{\lambda_e} 
&=& \lambda_e\dfrac{\beta e^4  \Sigma_\psi}{2^9 \pi^2 \gamma^4 M^2}\Big\{\cos{\theta}(1 + \alpha_\psi)(s_z + s'_z) +
+ \sqrt{1-\alpha_\psi^2} \nonumber \\
&\times & \sin \theta \Big[ \sin \Delta \Phi_{EM} (s_y s'_z + s_z s'_x) 
+ \cos \Delta \Phi_{EM} (s_x + s'_x) \Big]  \nonumber \\  
&+& f_\psi \sqrt{1+\alpha_\psi} \sin \theta \Big[ \sin \Delta \Phi_{M3} (s_z s'_x - s_x s'_z) \nonumber \\
&-&  \cos \Delta \Phi_{M3} (s_y - s'_y) \Big] \Big\}. 
\label{eq:CS_with_polarization}
\end{eqnarray}
 Appendix \ref{sec:appex} presents some details of derivation of these equations.  

Finally, the EDM of a baryon, $d$, is related to ED form-factor $F_3(s)$ via   
\begin{equation}\label{eq8}
    d = \dfrac{e}{2 M} F_3(0).
\end{equation}
To use this formula one needs to analytically continue complex-valued form-factor to the 
real-photon point.  We assume that the form-factor varies weakly as a function of $s$.

\section{Generation of pseudo-data}
\label{sec:pseudo-data}

 The goal of this work is to perform the sensitivity study of the EDM with the current and future tau-charm factories, i.e. to estimate the achievable precision of EDM measurements. To do so, we first generate a {\it pseudo-data} with the statistics of BESIII and STCF and then, estimate the uncertainties to the theoretical parameters of question.  
 Our probability distribution function depends on five kinematical variables: 
\[ \theta, \,  \theta_-, \, \phi_-, \, \theta_+, \, \phi_+\] 
in addition to the six theoretical parameters that can be determined by fitting to the data: 
\[ \alpha_\psi, \, \Delta \Phi_{EM}, \, \Delta \Phi_{M3}, \, f_\psi, \alpha_-, \, \alpha_+ .\]
In this work, we will also vary the polarization of the initial electron $\lambda_e$.

The probability distribution function for process $e^+ e^- \rightarrow  \Lambda \bar{\Lambda} \rightarrow p\pi^- \bar{p} \pi^+$ is defined as follows using (\ref{eq4}):
\begin{eqnarray}\label{eq3.1}
   \mathcal{W}(\theta, \theta_-, \phi_-, \theta_+, \phi_+;&& \hspace*{-0.5cm}\alpha_\psi, \Delta \Phi_{EM}, \Delta \Phi_{M3}, f_\psi, \alpha_-, \alpha_+; \lambda_e)  \nonumber\\
    &=& \dfrac{1}{\sigma (e^+ e^- \rightarrow  \Lambda \bar{\Lambda} \rightarrow p\pi^- \bar{p} \pi^+)}\frac{d \sigma (e^+ e^- \rightarrow  \Lambda \bar{\Lambda} \rightarrow p\pi^- \bar{p} \pi^+)}{d \cos \theta d\Omega_- d\Omega_+},
\end{eqnarray}
where we can calculate $\sigma (e^+ e^- \rightarrow  \Lambda \bar{\Lambda} \rightarrow p\pi^- \bar{p} \pi^+)$ by integrating Eq.~(\ref{eq4}) over the full region of kinematical angles. The similar procedure 
can be performed for the process $e^+ e^- \rightarrow \Lambda_c^+ \bar{\Lambda}_c^- \rightarrow \Lambda\pi^+ \bar{\Lambda} \pi^-$. Next, using this formula we generate a set of events, via hit-or-miss method, with predefined theoretical parameters as explained below. 

First, let us estimate the statistics available for the BESIII and STCF experiments. 
We start with the process of $e^+ e^- \rightarrow J/\psi \rightarrow \Lambda \bar{\Lambda} \rightarrow p\pi^- \bar{p} \pi^+$. It is known that for ten years of work, from 2009 to 2019, BESIII obtained 
$J/\psi$ statistics of approximately $10^{10}$ events \cite{BESIII:2021cxx}, while the 
STCF experiment plans to obtain $3.4 \times 10^{12}$ $J/\psi$ events per year~\cite{Achasov:2023gey}. 
Now, using the measured branching fraction \cite{ParticleDataGroup:2022pth} for the 
decays $J/\psi \rightarrow \Lambda \bar{\Lambda}$ and $\Lambda \rightarrow p \pi^-$, 
which are respectively ${\rm Br}(J/\psi \rightarrow \Lambda \bar{\Lambda}) = 1.89 \times 10^{-3}$ and 
${\rm Br}(\Lambda \rightarrow p \pi^-) = 0.64$,  and assuming the detection efficiency of $40\%$,  
we can obtain the  statistics of the process under consideration. 
For $e^+ e^- \rightarrow J/\psi \rightarrow \Lambda \bar{\Lambda} \rightarrow p\pi^- \bar{p} \pi^+$, 
the statistics that corresponds to ten years of observations in the BESIII experiment is 
$3.1 \times 10^6$ events~\cite{BESIII:2023rwv}, and for one year of observations in 
the STCF experiment we obtain  $10.6 \times 10^8$ events per year~\cite{Achasov:2023gey}. 
The assumed luminosities of the BESIII \cite{BESIII:2021cxx} and STCF \cite{Achasov:2023gey} 
experiments for the $e^+e^- \rightarrow J/\psi$ process are $2678$ pb$^{-1}$ and 1 ab$^{-1}$, respectively.

Under the similar condition, for the process $e^+ e^- \rightarrow \Lambda_c^+ \bar{\Lambda}_c^- \rightarrow \Lambda\pi^+ \bar{\Lambda} \pi^-$, we find that a sufficient statistics for a full angular 
analysis can be collected only by the STCF experiment: it is  planned to generate 
$5.6 \times 10^8$ pairs of  $\Lambda_c^+ \bar{\Lambda}_c^-$ per year \cite{Achasov:2023gey}. 
Using the known branching fraction \cite{ParticleDataGroup:2022pth},  
${\rm Br}(\Lambda_c^+ \rightarrow \Lambda \pi^+) = 1.3 \times 10^{-2}$, and 
assuming the  expected detection efficiency similar to the one from BESIII, $42\%$ \cite{BESIII:2015bjk}, 
we can obtain $4 \times 10^4$ events of the desired process per year. In the following, we 
will consider the statistics that will be collected over three years. The luminosity of 
the STCF \cite{Achasov:2023gey} experiment is assumed to be 1 ab$^{-1}$.

Taking into account all of the above, we generate pseudo-data and fit the theoretical parameters 
in the next section. In order to generate the pseudo-data, we need the inputs of the 6 theoretical 
parameters listed earlier. Apart from the two parameters related to EDM, $f_\psi$ and 
$\Delta \Phi_{M3}$, which are unknown, we have some information for the rest of the 
parameters from previous measurements done without EDM terms. 
\begin{table}[h]
\begin{center}
\begin{tabular}{c c c c}
\hline
$\sqrt{s}$, GeV & 4.6119 & 4.6819 & 4.9509 \\
\hline
$\alpha_\psi$ & -0.26 & 0.15 & 0.63 \\
\hline
\end{tabular}
\end{center}
\caption{\label{tab1}Dependence of $\alpha_\psi$ on $\sqrt{s}$ \cite{BESIII:2023rwv}.}
\end{table}

\begin{figure}[htb]
\begin{center}
\includegraphics[width=0.45\linewidth]{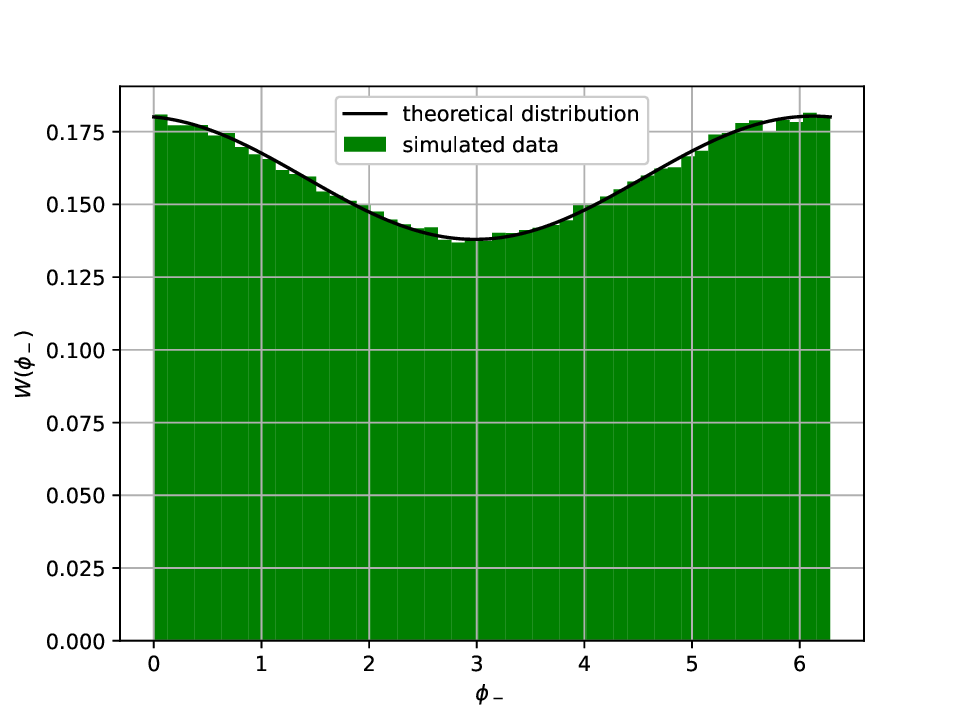}
\includegraphics[width=0.45\linewidth]{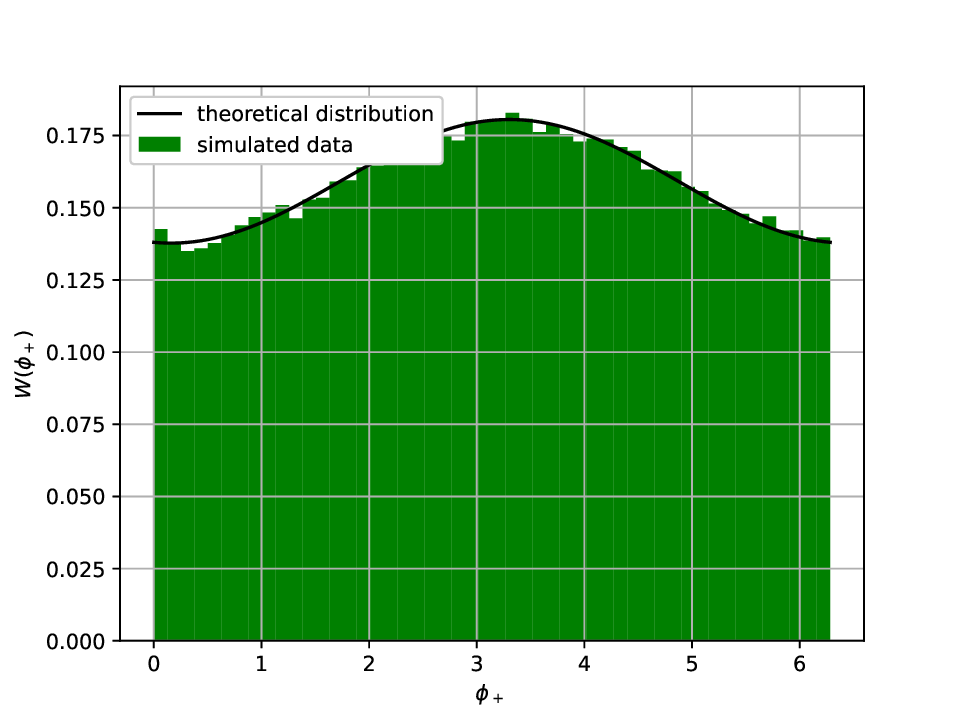}\\
\includegraphics[width=0.45\linewidth]{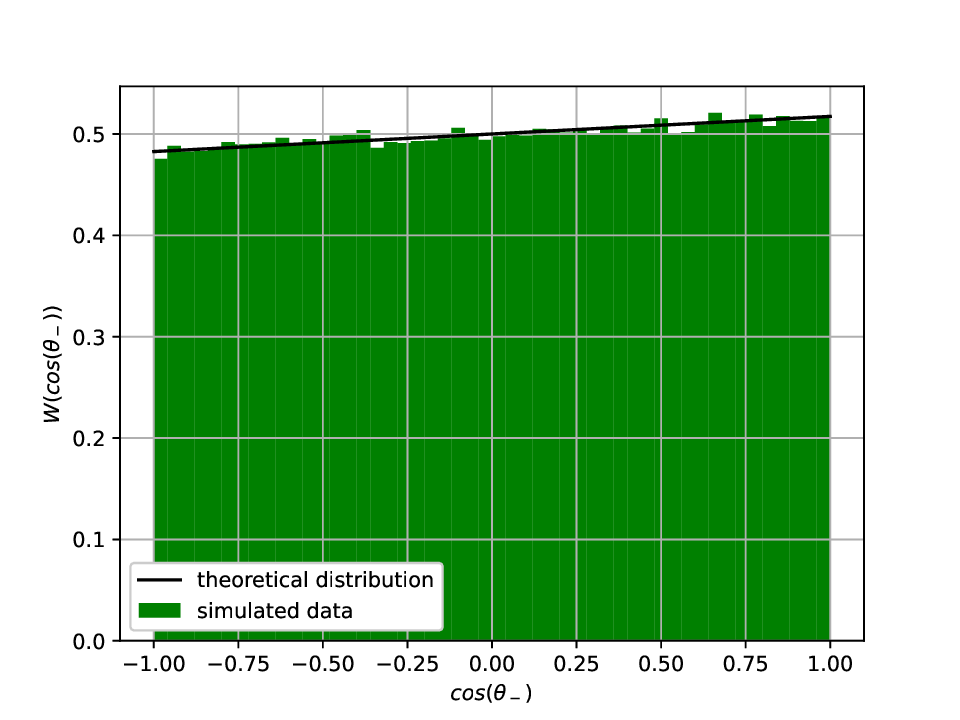}
\includegraphics[width=0.45\linewidth]{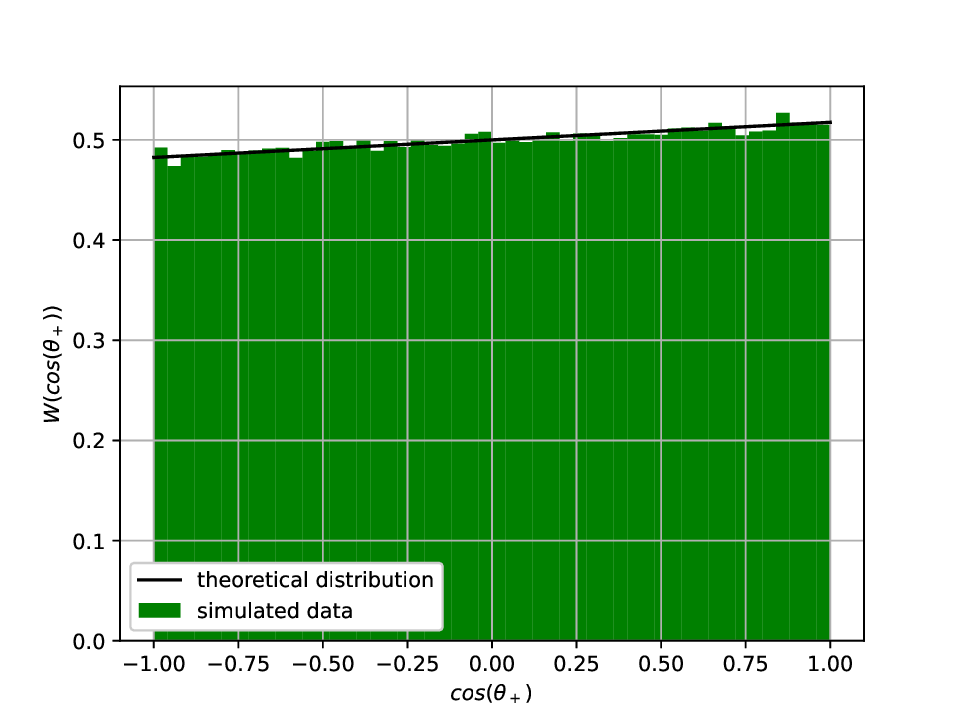}\\
\includegraphics[width=0.45\linewidth]{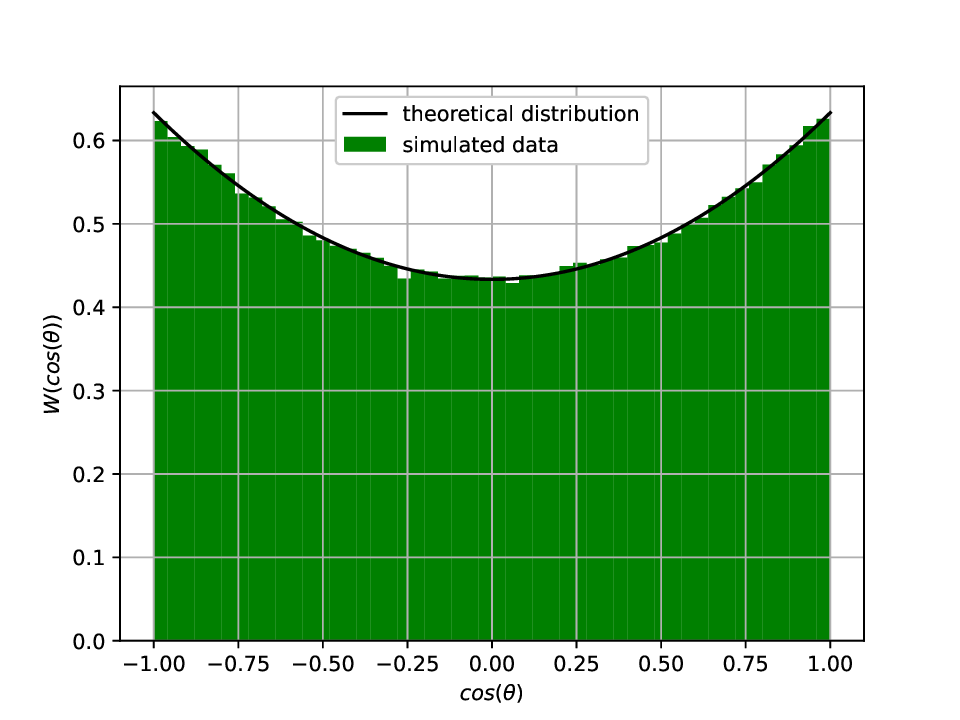}
\caption{Comparison of the theoretical one-dimensional distribution with the distribution obtained by Monte Carlo simulation: top-left $\mathcal{W}(\phi_-)$, top-right $\mathcal{W}(\phi_+)$, middle-left $\mathcal{W}(\cos\theta_-)$, middle-right $\mathcal{W}(\cos\theta_+)$, bottom $\mathcal{W}(\cos\theta)$.    
Number of events is $5 \times 10^5$.} 
\label{fig2} 
\end{center}
\end{figure}

For the process $e^+ e^- \rightarrow  \Lambda \bar{\Lambda} \rightarrow p\pi^- \bar{p} \pi^+$ we will use the following values known from experiments: $\alpha_\psi = 0.46$, and this $\alpha_\psi$ value corresponds 
to the $J/\psi$ energy of the resonance, $\Delta \Phi_{EM} = 0.74$, $\alpha_- = 0.75$, 
$\alpha_+ = -0.758$~\cite{BESIII:2018cnd}. 

For process  $e^+ e^- \rightarrow  \Lambda_c^+ \bar{\Lambda}_c^- \rightarrow \Lambda \pi^+ \bar{\Lambda} \pi^-$, we also know several parameters from experiments that will be used in generating pseudo-statistics: 
$\alpha = - 0.84$, $\bar{\alpha} = 0.97$~\cite{ParticleDataGroup:2022pth}. 
Also note that since $\alpha_\psi$ depends directly on the moduli of the form factors, which in turn depend on $s$, we can look at this process for different energies of the center of mass. Thus in Table~\ref{tab1} values of  $\alpha_\psi$ are shown at different cms energies \cite{BESIII:2023rwv}. For the unknown parameters $f_\psi$ and $\Delta \Phi_{M3}$, these will be varied to estimate the achievable precisions for each value.

Now, as examples, let us show how the  pseudo-data look like for the 
process $e^+ e^- \rightarrow \Lambda \bar{\Lambda} \rightarrow p\pi^- \bar{p} \pi^+$. The following parameter values are used to construct these model data: $\alpha_\psi = 0.461$, $\Delta \Phi_{EM} = 0.74$, 
$\Delta \Phi_{M3} = 0.5$, $f_\psi = 0.1$, $\alpha_- = 0.75$, $\alpha_+ = -0.758$ and $\lambda_e = 0.5$. The number of events is $5 \times 10^5$. 
Figs.~ \ref{fig2} shows one-dimensional histograms of the event distribution for each of the five angles. The theoretical curves, against which the behavior of the experimental distribution is compared, are obtained by 
integration of the distribution (\ref{eq3.1}) over all angles except the one for which 
the distribution is plotted.

\section{Results}
\label{sec:results}

 Having the generated pseudo-data, we use the maximum likelihood method to obtain the best fit values of the theoretical parameters, their errors and correlation matrices. 

Let us start with the $e^+ e^- \rightarrow \Lambda \bar{\Lambda} \rightarrow p\pi^- \bar{p} \pi^+$ process. 
We use  the same parameter values and the same sample size for which the distributions in Figs.~\ref{fig2} are obtained, and determine the standard deviation and correlation matrix.  As a result, we obtain the following fitted parameters: 
\begin{eqnarray}
    &&\hat{\alpha}_\psi = 0.4597,\quad \Delta\hat{\Phi}_{EM} = 0.7490, \quad \hat{\alpha}_- = 0.7463,\quad \hat{\alpha}_+ = -0.7669, \nonumber\\ 
    &&\hat{f}_\psi = 0.0946, \quad \Delta\hat{\Phi}_{M3} = 0.5576,
		\label{eq:fitted_parameters}
\end{eqnarray}
and the corresponding standard deviations:
\begin{eqnarray}
    &&\sigma_\psi = 0.0042,\quad \sigma_\Phi = 0.0071, \quad\sigma_- = 0.0038,\quad \sigma_+ = 0.0039, \nonumber\\
    &&\sigma_f = 0.0035,\quad \sigma_{M3}= 0.0496.
		\label{eq:standard_deviations}
\end{eqnarray}

The correlation matrix can also be obtained: 
\begin{equation}
\rho_{ij} = 
    \begin{pmatrix}
    1 & 0.0081 & -0.0488 & 0.0331 & 0.0526 & 0.0125 \\
    0.0081  & 1 & -0.0388 & 0.0240 & -0.0164 & -0.0294 \\
    -0.0488  & -0.0388 & 1 & 0.4214 & -0.0394 & -0.0042\\
    0.0331 & 0.0240 & 0.4214 & 1 & 0.0557 & 0.0035 \\
    0.0526 & -0.0164 & -0.0394 & 0.0557 & 1 & -0.1436\\
    0.0125 & -0.0294 & -0.0042 & 0.0035 & -0.1436 & 1 
    \end{pmatrix},
		\label{eq:correlation_matrix}
\end{equation}
where we observe a large correlation between the parameters $\alpha_+$ and $\alpha_-$, which, apparently, is due to the proximity of the modules of these parameters at the initial choice of their values when generating the pseudo-statistics.

In this example, the input values for the two ED form-factors are fixed to be $f_\psi=0.3$ and $\Delta \Phi_{M3}=0.5$. Next, we will investigate the dependence of the sensitivity on these parameters. 

\begin{figure}[htb]
\begin{center}
\includegraphics[width=0.70\linewidth]{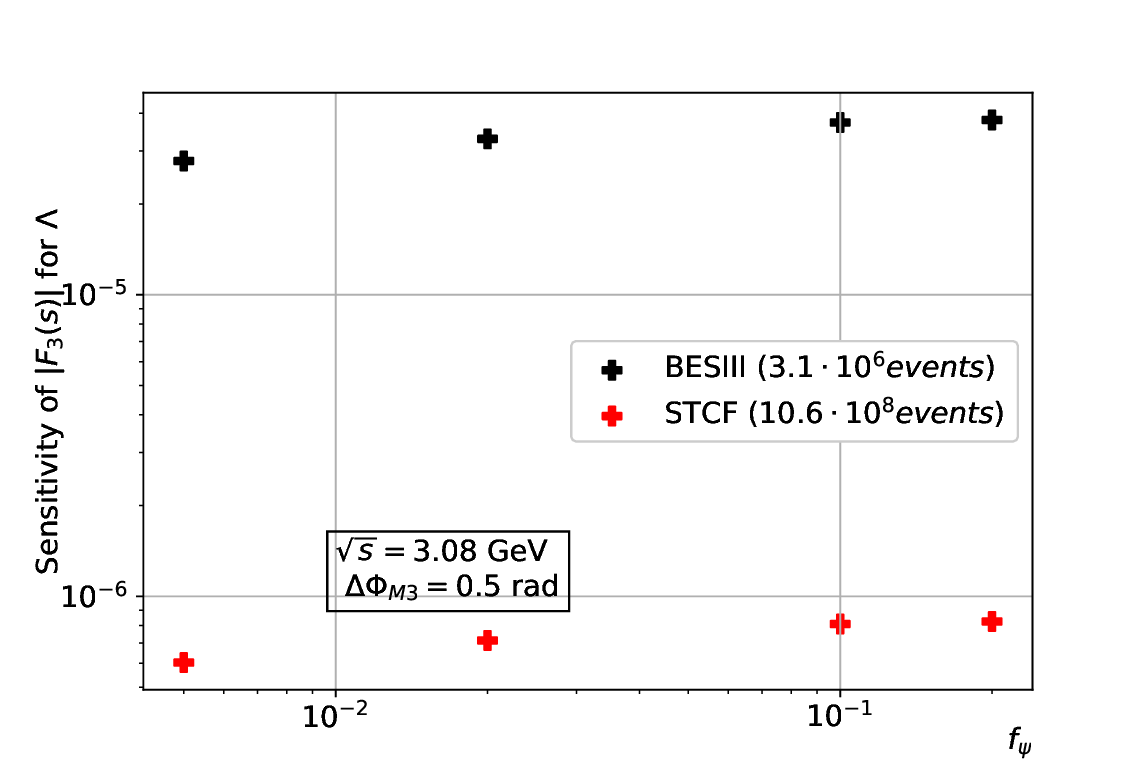}
\includegraphics[width=0.65\linewidth]{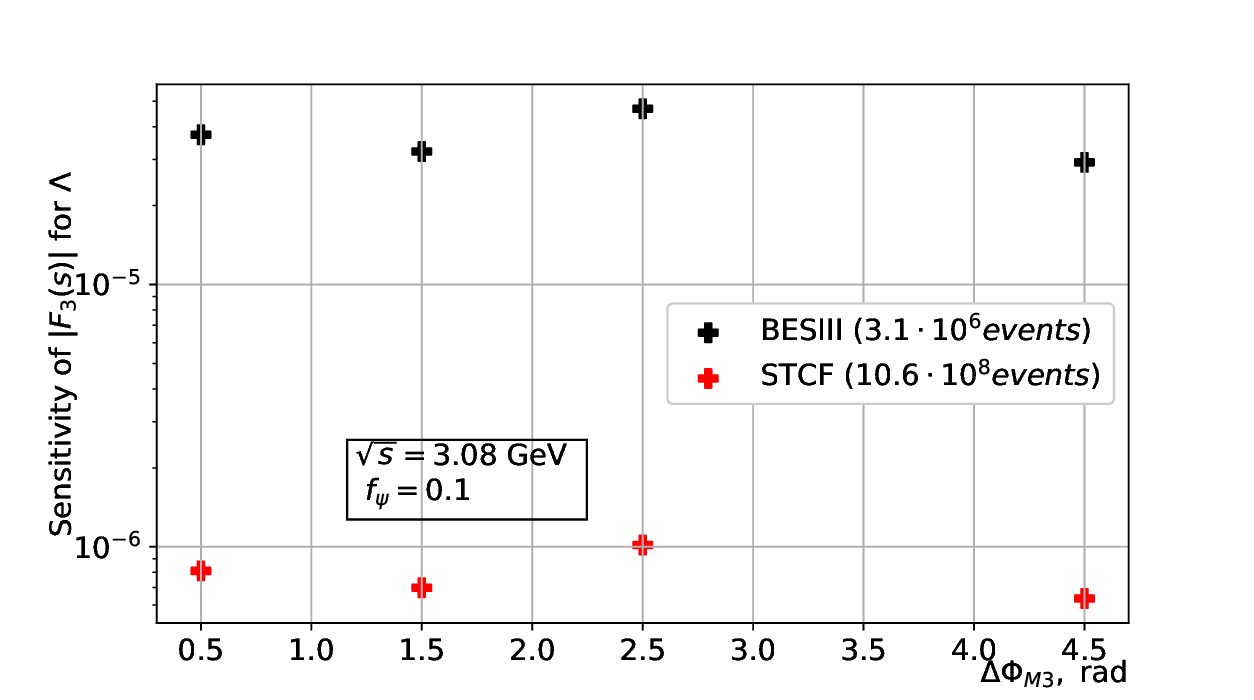}
\includegraphics[width=0.63\linewidth]{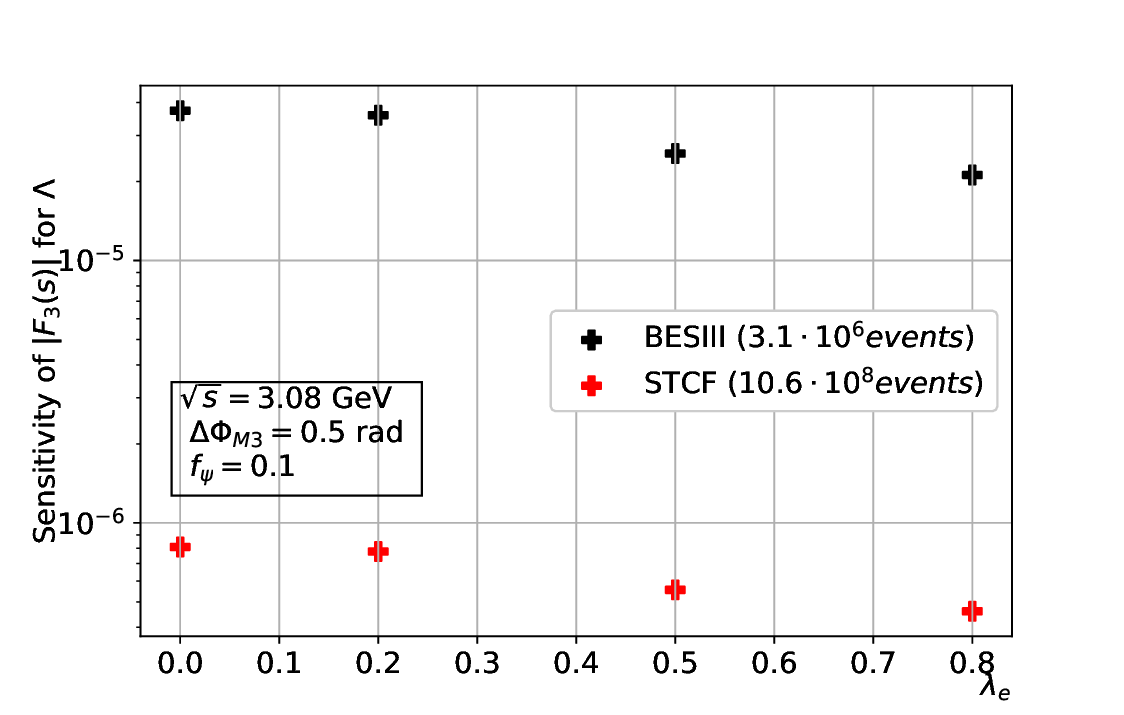}
\caption{Dependence of the sensitivity of the $\Lambda$ ED form-factor measurement on different values 
of $f_\psi$ (top), $\Delta \Phi_{M3}$ (middle) and $\lambda_e$ (bottom). Other unknown parameters are fixed, and their values are shown in the plot. There is no polarization of the initial electron in the top and middle plots. } 
\label{fig:3} 
\end{center}
\end{figure}

In Fig.~\ref{fig:3} top, we show our result of sensitivity dependence of the ED form-factor measurements on the input parameters $f_\psi$.The red points represent the statistics of BESIII experiment and blue points are for STCF. Firstly, we can observe a general tendency that the error is inversely proportional to the square root of the statistics. The dependence of sensitivity on $f_\psi$ is very small. 

Similarly, from Fig.~\ref{fig:3} middle, it is seen that the measurement error of the ED form-factor slightly oscillates
with variation of the phase difference $\Delta \Phi_{M3}$.  

As for the dependence on the polarization of the initial electron -- with 
increasing the polarization there is a tendency of reducing the ED form-factor measurement error, i.e., of increasing 
the sensitivity, as shown in Fig.~\ref{fig:3} bottom. 

In general, one has the relation (\ref{eq8}) between the form-factor $F_3 (s)$ and EDM at $s=0$, 
thus from the above results we can  estimate approximately the sensitivity of EDM.  
Then we find that the error of the $\Lambda$-baryon EDM measurement for the BESIII experiment is
\begin{equation}
    \delta d_\Lambda^{(BESIII)} \sim 0.5 \times 10^{-18} \, e \,{\rm cm},
		\label{eq:EDM_Lambda_BESIII}
\end{equation}
and for the STCF experiment:
\begin{equation}
    \delta d_\Lambda^{(STCF)} \sim 1.5 \times 10^{-20} \,  e \,{\rm cm}.
		\label{eq:EDM_Lambda_STCF}
\end{equation}

\begin{figure}[htb]
\begin{center}
\centerline{\includegraphics[scale=0.32]{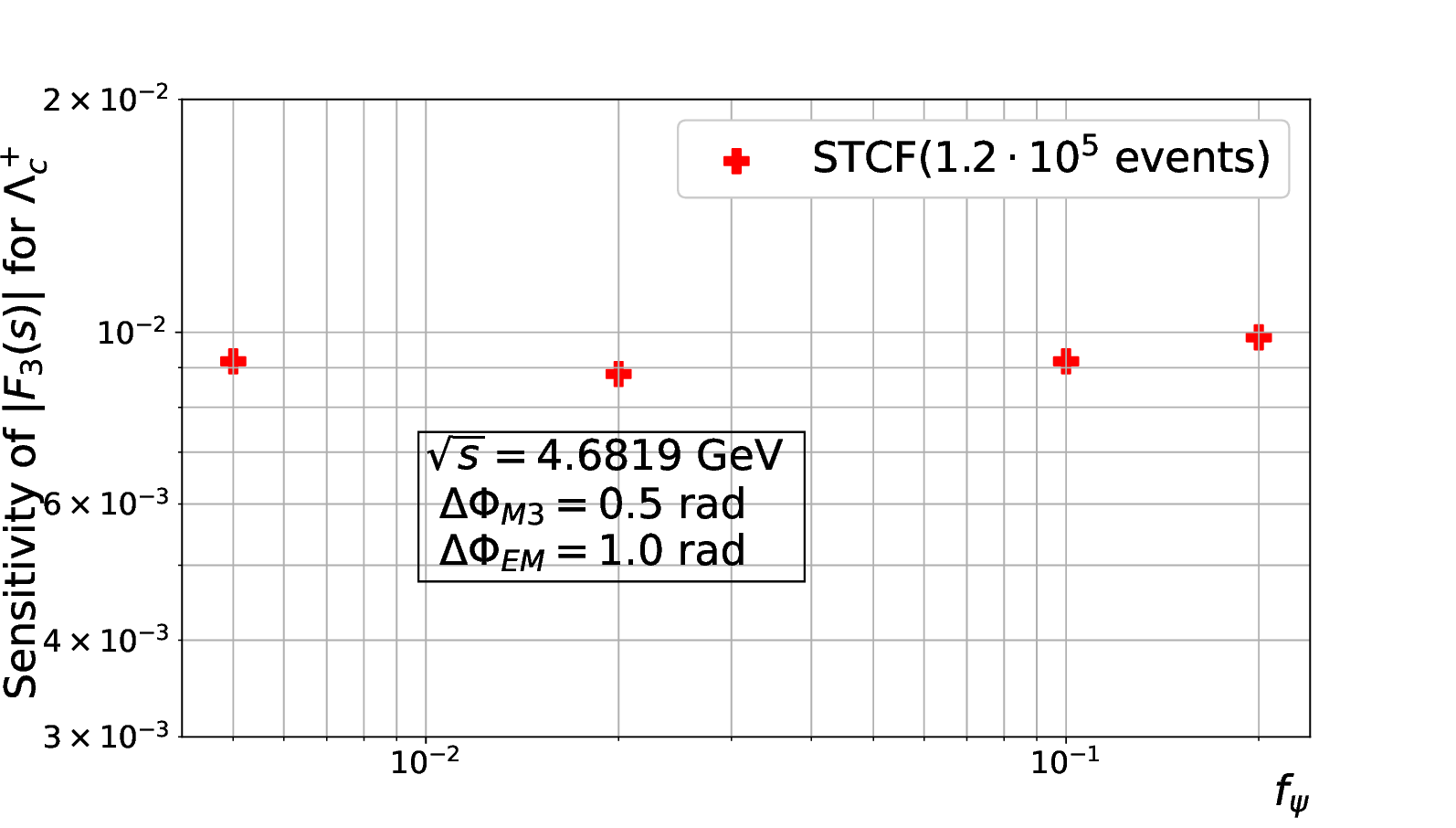}}
\centerline{\includegraphics[scale=0.35]{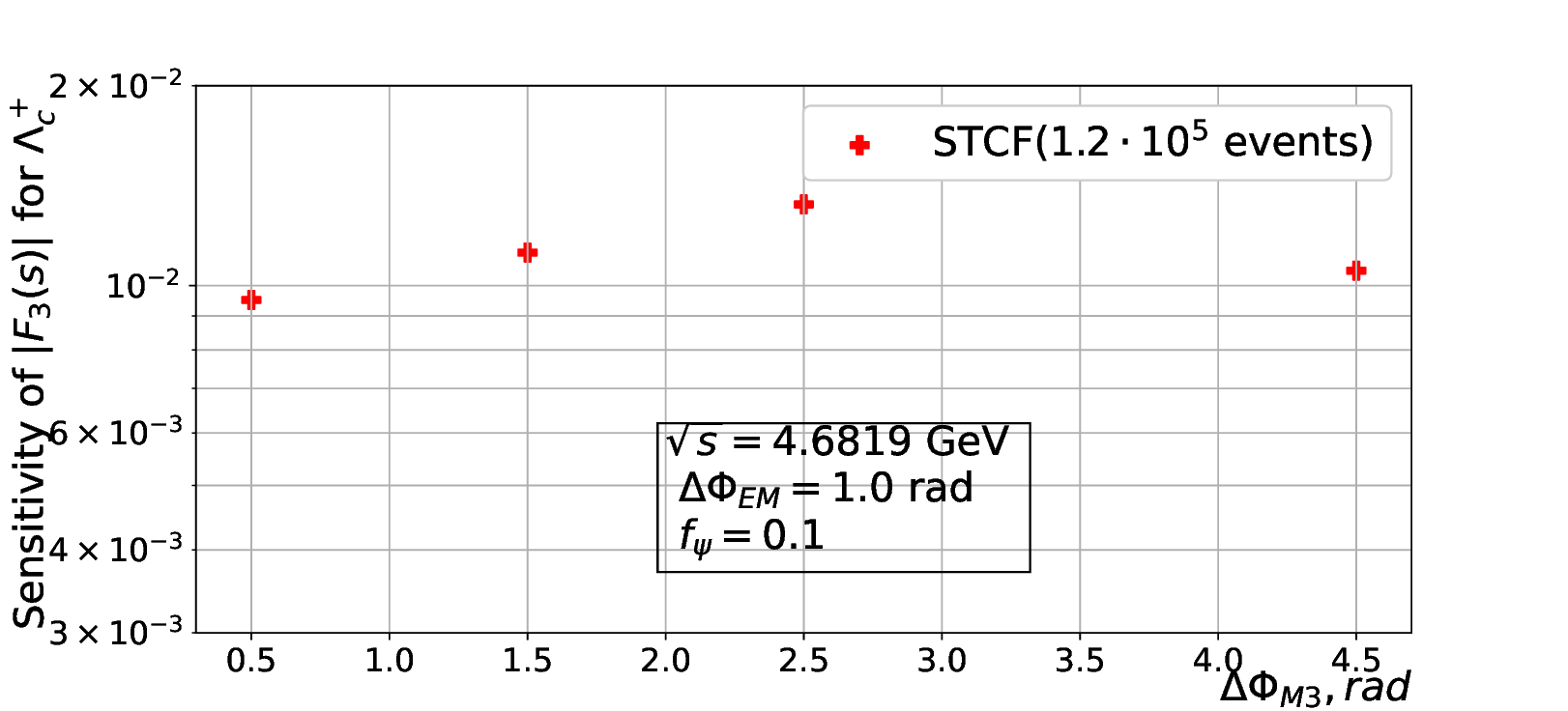}}
\centerline{\includegraphics[scale=0.36]{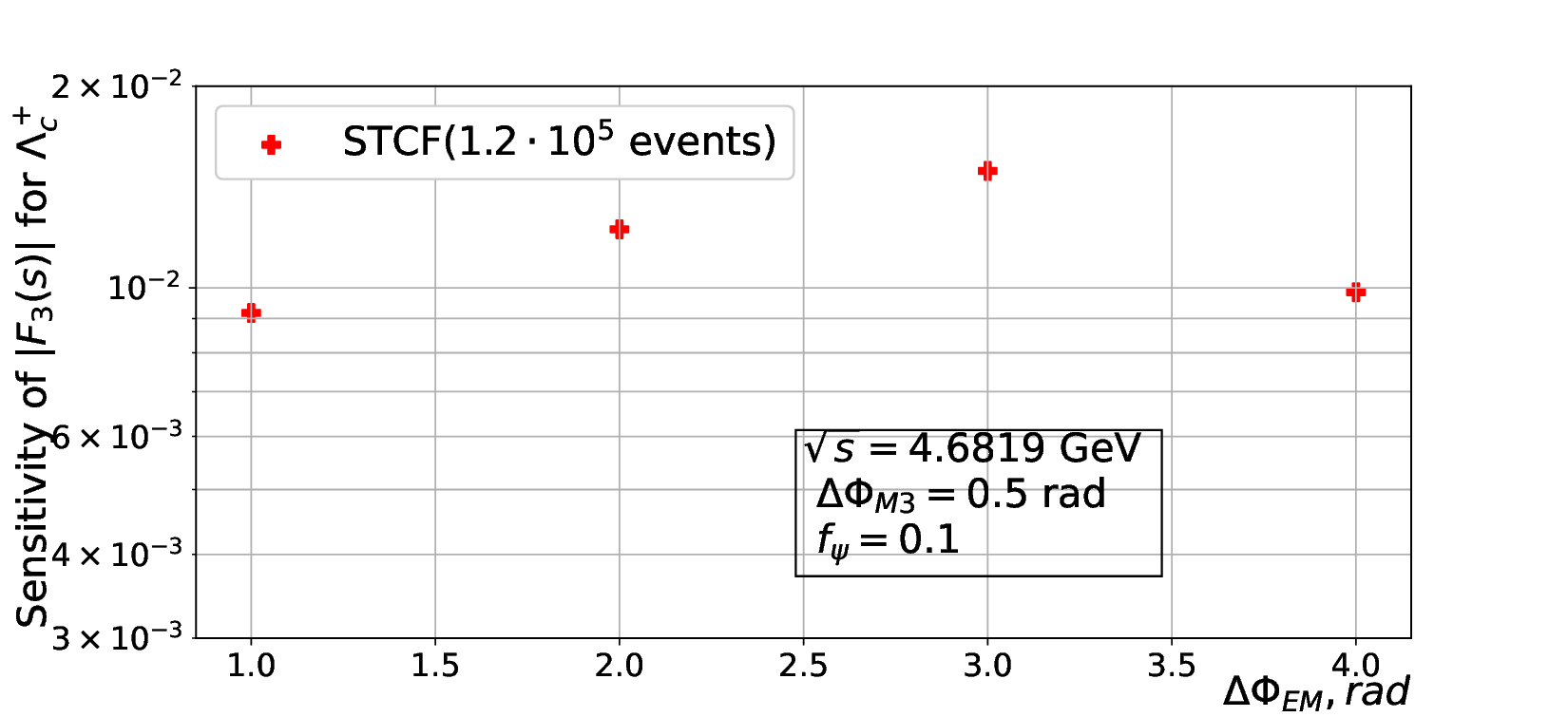}}
\caption{Dependence of the sensitivity of the $\Lambda_c^+$ ED form-factor measurement, from top to bottom,  
on $f_\psi$, $\Delta \Phi_{M3}$ and $\Delta \Phi_{EM}$. Other unknown parameters are fixed, and their values are shown in the plot. The polarization of the initial electron is absent.} %
\label{fig:41} 
\end{center}
\end{figure}

\begin{figure}[htb]
\begin{center}
\centerline{\includegraphics[scale=0.4]{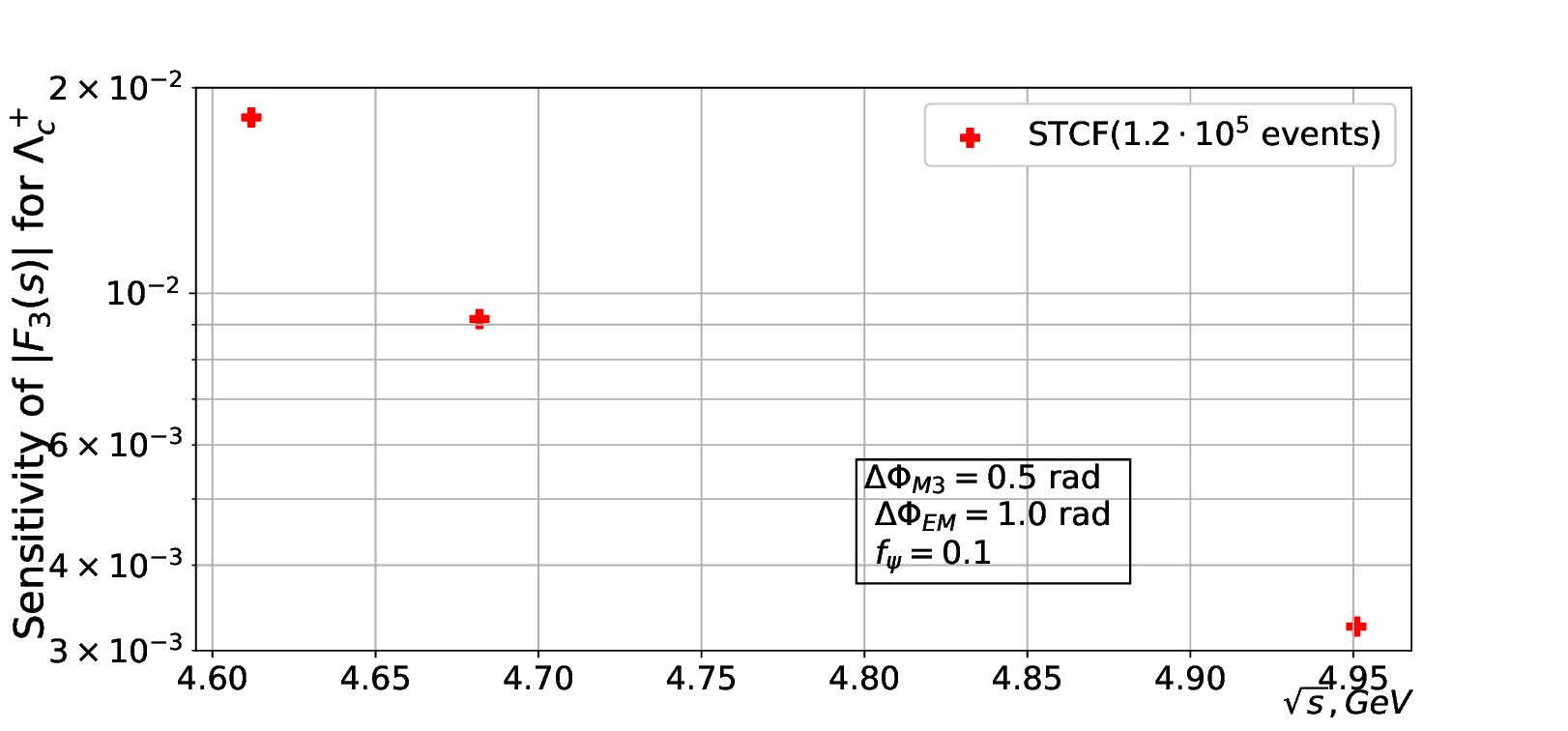}}
\centerline{\includegraphics[scale=0.39]{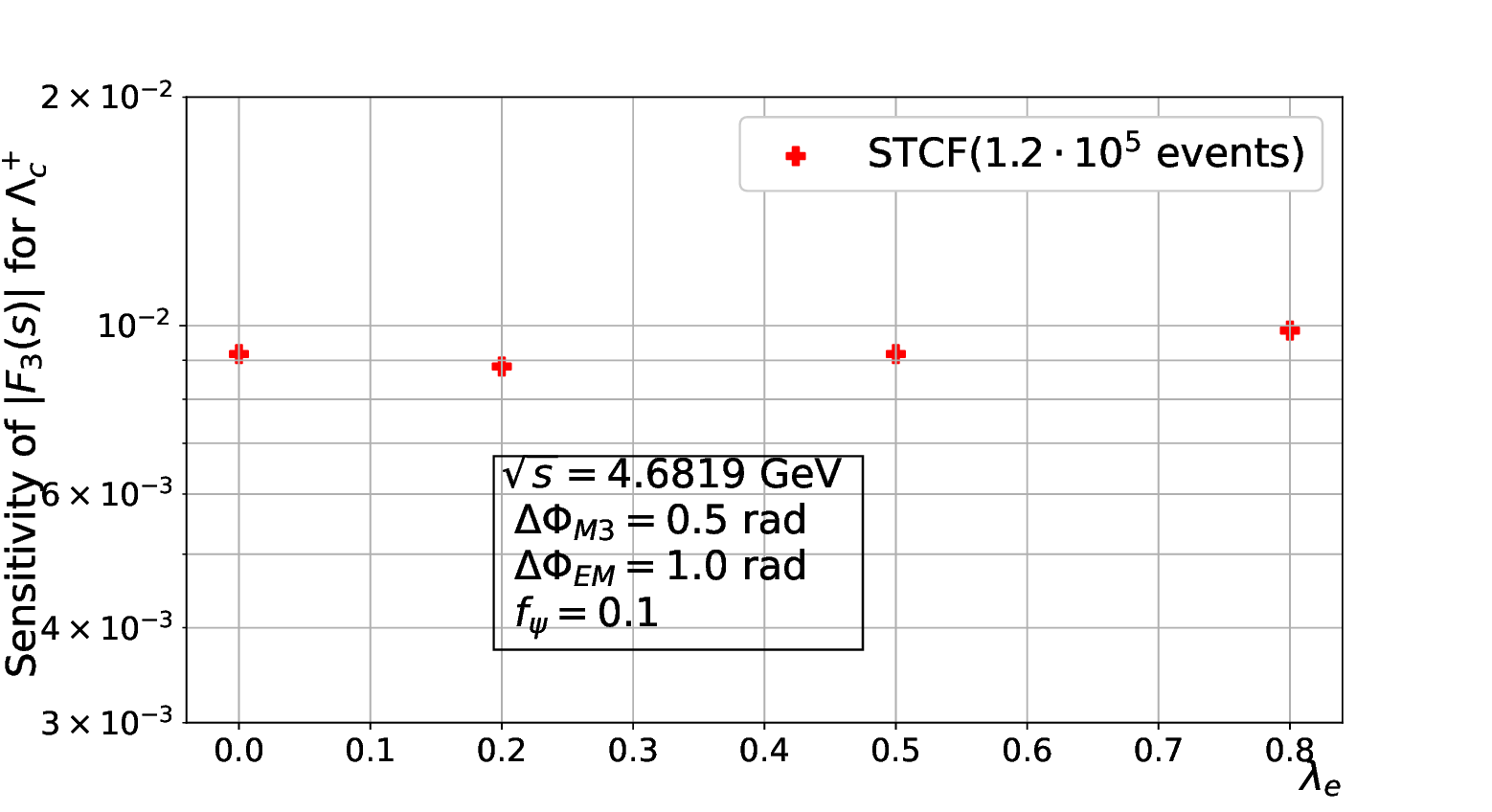}}
\caption{Dependence of the sensitivity of the $\Lambda_c^+$ ED form-factor measurement, from top to bottom,  on $\sqrt{s}$ and $\lambda_e$.  Other unknown parameters are fixed, and their values are shown in the plot.} 
\label{fig:5} 
\end{center}
\end{figure}

The analysis of the process $e^+ e^- \rightarrow \Lambda_c^+ \bar{\Lambda}_c^- \rightarrow \Lambda\pi^+ \bar{\Lambda} \pi^-$ is carried out in the same way as for $e^+ e^- \rightarrow \Lambda \bar{\Lambda} \rightarrow p\pi^- \bar{p} \pi^+$. Compared to $\Lambda$ EDM, we have more unknown parameters, such as $\alpha_\psi$ and $\Delta \Phi_{EM}$. 
We investigate the dependence of the sensitivity on these input parameters in Fig.~\ref{fig:41}. We find that the sensitivity of the ED form-factor measurement of $\Lambda_c^{+}$ baryon is stable, less than 50\%, to the variation of these input parameters. 
On the other hand, a clear increase in sensitivity (decrease in measurement error) is observed, by an order of magnitude, with increasing  energy as seen in  Fig.~\ref{fig:5} (top). 
In Fig.~\ref{fig:5} (bottom), one can see that the ED form-factor measurement error of the $\Lambda_c^+$ baryon does not depend much on electron  polarization.  

Next, as in the case of $\Lambda$, now we can give an estimate for the sensitivity of the EDM of $\Lambda_c^+$. 
Summarizing all of the above, we can conclude that the  error  the EDM measurement for the $\Lambda_c^+$ baryon in the STCF experiment is:
\begin{equation}
    \delta d_{\Lambda_c^+}^{(STCF)} \sim 0.8 \times 10^{-16} \, e \,{\rm cm}.
		\label{eq:EDM_Lambda_c_STCF}
\end{equation}


\section{Conclusions}
\label{sec:conclusions}

In this paper, we have investigated a future measurement of the EDM of $\Lambda$ and $\Lambda_c$ baryons at tau-charm factories, namely BESIII and STCF experiments. We first derived the theoretical framework to exploit the  full angular distributions of $e^+ e^- \rightarrow \Lambda \bar{\Lambda} \rightarrow p\pi^- \bar{p} \pi^+$ and $e^+ e^- \rightarrow \Lambda_c^+ \bar{\Lambda}_c^- \rightarrow \Lambda\pi^+ \bar{\Lambda} \pi^-$ processes, including the EDM contribution. As the STCF experiment may be able to produce polarized beams, we have also introduced polarization parameter to investigate its impact on measuring the EDM. The obtained result was used for Monte Carlo study to demonstrate the sensitivity of the future experiments to the EDM. We first have generated pseudo-data corresponding to the statistics of the BESIII and STCF experiments, in order to produce the full kinematical distributions. 
Then, by using the maximum likelihood method, we have shown that all the theoretical parameters, 6 for the former process and 8 for the latter, can be fitted by using the full angular distributions. 
We have also investigated the dependence of the sensitivity of the EDM measurement on the input parameters. We found that the sensitivity does not depend strongly on the size of EDM that we have introduced. We have found that the polarized beam can help to increase the sensitivity to the EDM measurement. In case of $\Lambda_c^+$, we also found that the sensitivity increases with a higher beam energy. 
Finally, we obtained sensitivity to the EDM for different experiments as: 
$\delta d_\Lambda^{(BESIII)} \sim 0.5 \times 10^{-18} \, e \,{\rm cm}$ for the $\Lambda$-baryon in the BESIII experiment, $\delta d_\Lambda^{(STCF)} \sim 1.5 \times 10^{-20} \, e \,{\rm cm}$ in the STCF experiment; $\delta d_{\Lambda_c^+}^{(STCF)} \sim 0.8 \times 10^{-16} \, e \,{\rm cm}$ for the $\Lambda_c^+$ baryon. It appeared that the latter number is very close to the expected sensitivity of the proposed new experiment~\cite{Aiola:2020yam} to measure the EDM of $\Lambda_c^+$  using bent crystals. Thus, we emphasize that the large production rate of the $\Lambda_c^+$ at STCF will play an important role in this field of research in the near future.

\vskip0.5cm
\section*{ACKNOWLEDGMENTS}

A.Yu.K. is grateful to University Paris-Saclay for financial support through the grant Erasmus+ ICM.  
He also thanks Jagiellonian University in Krakow, Poland, for support.  
R.T.O. acknowledges support from the Agence Universitaire de la Francophonie under the call for proposals in the discipline “Sciences de la matière - Physics” and for collaborating with teams from the PHE  and the Theory Departments of IJCLab. The authors would also like to thank S. Barsuk for his constant supports throughout this work.

\vskip0.5cm
\section*{Note added in proof}
While we were finalizing this article, a new result of the BESIII experiment appeared~\cite{BESIII:2025vxm}, in which one of the measurements that we propose in this article, the EDM of $\Lambda$ baryon, is performed. The obtained limit is similar to our estimate.

\appendix

\vspace{1cm}
\section{Matrix element squared in terms of spin-correlation matrix}
\label{sec:appex}

The squared  matrix element for the electron-positron pair annihilation 
into a polarized pair $ \Lambda \bar{\Lambda}$ (or 
$\Lambda_c^+ \bar{\Lambda}_c^-$) and polarized electron can be represented in the form:
\begin{equation}
\overline{|\mathcal{M}|^2} = R_{00} + \sum_{i=1}^3 R_{i0} s_i + 
\sum_{j=1}^3 R_{0j} s_j' + \sum_{i,j=1}^3 R_{ij} s_i s_j',
\end{equation}
where $s_i$ and $s_j'$ are the polarization vectors of $\Lambda$ and $\bar{\Lambda}$ in their rest frames, respectively. The elements of the spin-correlation matrix $R_{00}$, $R_{i0}$, $R_{0j}$, and $R_{ij}$ are obtained by direct calculation:
\begin{equation}
(1 \times  1'): \quad R_{00} = \dfrac{e^4}{4 \gamma^2} \bigl[ (\gamma^2 - 1)\gamma^2 |F_3|^2 \sin^2\theta + |G_E|^2\sin^2\theta + \gamma^2 |G_M|^2(\cos^2 \theta + 1)\bigr], 
\end{equation}
\begin{equation}
(s_1 \times s_1'): \quad R_{11} = -\dfrac{e^4 \sin^2\theta}{4 \gamma^2} \bigl[ (\gamma^2 - 1)\gamma^2 |F_3|^2 - |G_E|^2 - \gamma^2 |G_M|^2 \bigr], 
\end{equation}
\begin{equation}
(s_1 \times s_2'): \quad R_{12} = - R_{21} = \dfrac{e^4 \beta \sin^2\theta}{2} {\rm Re}(G_E F_3^*),
\end{equation}
\begin{equation}
(s_2 \times s_2'): \quad R_{22} = -\dfrac{e^4 \sin^2\theta}{4 \gamma^2} \bigl[ (\gamma^2 - 1) \gamma^2 |F_3|^2 - |G_E|^2 + \gamma^2 |G_M|^2 \bigr], 
\end{equation}
\begin{equation}
(s_1 \times s_3'): \quad R_{13} = \dfrac{e^4 \sin \theta}{2 \gamma} \bigl[ {\rm Re}(G_M G_E^*)\cos\theta 
-  \lambda_e \beta \gamma^2 {\rm Im}(G_M F_3^*) \bigr],
\end{equation}
\begin{equation}
(s_3 \times s_1'): \quad R_{31}  = \dfrac{e^4 \sin \theta}{2 \gamma} \bigl[ {\rm Re}(G_M G_E^*)\cos\theta 
+ \lambda_e \beta \gamma^2 {\rm Im}(G_M F_3^*) \bigr],
\end{equation}
\begin{equation}
(s_2 \times s_3'): \quad R_{23} = -\dfrac{e^4 \sin \theta}{2\gamma} \bigl[ \beta \gamma^2 {\rm Re}(G_M F_3^*)\cos \theta + \lambda_e  {\rm Im}(G_M G_E^*)\bigr],
\end{equation}
\begin{equation}
(s_3 \times s_2'): \quad R_{32} = \dfrac{e^4 \sin \theta}{2\gamma} \bigl[ \beta \gamma^2 {\rm Re}(G_M F_3^*)\cos \theta - \lambda_e {\rm Im}(G_M G_E^*)  \bigr],
\end{equation}
\begin{equation}
(s_3 \times s_3'): \quad R_{33} = \dfrac{e^4}{8 \gamma^2} \bigl\{ \gamma^2 |G_M|^2 (\cos 2 \theta + 3) - 2\sin^2\theta \bigl[ (\gamma^2 - 1) \gamma^2 |F_3|^2 + |G_E|^2 \bigr]  \bigr\}, 
\end{equation}
\begin{equation}
(s_1 \times 1'): \quad R_{10} = -\dfrac{e^4 \sin \theta}{2 \gamma} \bigl[ \beta \gamma^2 {\rm Im}(G_M F_3^*) \cos\theta - \lambda_e {\rm Re}(G_E G_M^*) \bigr],
\end{equation}
\begin{equation}
(1 \times s_1'): \quad R_{01} = \dfrac{e^4 \sin \theta}{2 \gamma} \bigl[ \beta \gamma^2 {\rm Im}(G_M F_3^*) \cos\theta + \lambda_e {\rm Re}(G_E G_M^*) \bigr],
\end{equation}
\begin{equation}
(s_2 \times 1'): \quad R_{20} = - \dfrac{e^4 \sin \theta}{2 \gamma} \bigl[{\rm Im}(G_M G_E^*) \cos \theta 
+ \lambda_e \beta \gamma^2 {\rm Re}(G_M F_3^*) \bigr],
\end{equation}
\begin{equation}
(1 \times s_2'): \quad R_{02} = - \dfrac{e^4 \sin \theta}{2 \gamma} \bigl[{\rm Im}(G_M G_E^*) \cos \theta 
- \lambda_e \beta \gamma^2 {\rm Re}(G_M F_3^*) \bigr],
\end{equation}
\begin{equation}
(s_3 \times 1'): \quad R_{30} = \dfrac{e^4}{2}\bigl[ \beta  {\rm Im}(G_E F_3^*) \sin^2 \theta
+  \lambda_e |G_M|^2 \cos \theta \bigr],
\end{equation}
\begin{equation}
(1 \times s_3'): \quad R_{03} = -\dfrac{e^4}{2}\bigl[\beta  {\rm Im}(G_E F_3^*) \sin^2 \theta
- \lambda_e |G_M|^2 \cos \theta \bigr].
\end{equation}
Here $\lambda_e$ denotes degree of the electron polarization, $\gamma=\sqrt{s}/(2M)$ and 
$\beta=\sqrt{1- \gamma^{-2}}$.


\vspace{1cm}
\bibliographystyle{unsrt}
\bibliography{EDM}

\end{document}